\documentclass{elsart}
\usepackage{natbib}
\usepackage{graphicx}
\begin{document}
\runauthor{Y.~Aritomo and M.~Ohta}
\begin{frontmatter}
\title{Trajectory analysis for fusion path in superheavy-mass region}
\author[Dubna]{Yoshihiro Aritomo}
\author[Kobe]{Masahisa Ohta}

\address[Dubna] {Flerov Laboratory of Nuclear
Reactions, JINR, Dubna, Russia}
\address[Kobe]{Department of Physics, Konan University, 8-9-1
  Okamoto, Kobe, Japan}

\begin{abstract}

We propose an effective method for the precise investigation of
the fusion-fission mechanism in the superheavy-mass region, using
the fluctuation-dissipation model. The trajectory calculation with
friction is performed in the nuclear deformation space using the
Langevin equation. In the reaction $^{48}$Ca+$^{244}$Pu, the
trajectories are classified into the fusion-fission process, the
quasi-fission process and the deep quasi-fission process. By
analyzing the time evolution of each trajectory, the mechanism of
each process is clearly revealed, i.e., it is explained why a
trajectory takes a characteristic path in this model. We discuss,
in particular, the condition under which the fusion path is
followed, which is crucial in the discussion of the possibility of
synthesizing superheavy elements.

\end{abstract}
\end{frontmatter}




\section{Introduction}

Experiments on the synthesis of superheavy elements using heavy
ion collisions have recently been successful in finding several
new elements, and the known area in the nuclear chart is
approaching the 'Island of Stability' step by step
\cite{ogan991,ogan01,ogan04,hofm00,mori03,mori04}.

In theoretical work, the dynamical aspects of the fusion-fission
mechanism have been investigated as a time evolution of the
nuclear shape with friction. In the 1980's, for the heavy ion
fusion reaction, the trajectory calculation in the nuclear
deformation space using a stochastic equation was performed by
Swiatecki et al \cite{swia81,bloc86}. They investigated the fusion
process by using the mean trajectory and discussed a dynamical
hindrance. As the degree of freedom of the nuclear shape, the
center of mass distance, mass asymmetry and neck diameter were
employed.

In the superheavy-mass region, however, even when substantial
extra push energy is supplied, it is very difficult for the mean
trajectory to reach the spherical region due to the strong Coulomb
repulsion force and strong dissipation force. Only the fluctuation
around the mean trajectory can reach the spherical region.
Therefore, in order to estimate the fusion probability, we must
introduce the dynamical model based on fluctuation-dissipation
theory. We have developed the model to apply the dynamical process
in the superheavy-mass region \cite{ari97}.

In the superheavy-mass region, due to the strong Coulomb repulsion
force, the nuclear shape is easily deformed in the fusion-fission
process. Therefore, it is extremely important to take into account
the deformation of the fragments in the trajectory calculation
\cite{ari04}. Consequently, as nuclear shape parameters, we employ
the center of mass distance, deformation of fragments and mass
asymmetry in the framework of two-center parameterization
\cite{maru72,sato78}.

Moreover, although in their previous calculation Swiatecki et al.
\cite{swia81,bloc86} used only the liquid drop model potential
energy surface, in the superheavy-mass region, the shell
correction energy plays a very important role
\cite{armb99,moll97}. It influences not only the fission process
but also the fusion process. In our model, the shell correction
energy is taken into account in the dynamical calculation.

To estimate the fusion probability precisely, we analyzed the time
evolution of the trajectory in the deformation space and
investigated whether the trajectory enters the fusion box, which
is a limited area in which the deformation space corresponds to
spherical nuclei \cite{ari04}. In our previous work \cite{ari04},
we have compared our calculation with experimental data. In the
reaction $^{48}$Ca+$^{244}$Pu, we showed the calculations of the
mass distribution of fission fragments agree with the experimental
data. Also the experimental data of the cross section derived by
counting mass symmetric fission events were reproduced by our
model. By analyzing the trajectories, we classified them into the
fusion-fission process (FF), the quasi-fission process (QF) and
the deep quasi-fission process (DQF) in the reaction \cite{ari04}.

On the basis of the results of our previous studies, we
investigate the essential factors for determining the trajectory's
behavior. The main purpose of the present work is to clarify the
mechanism of whole fusion-fission dynamics by analyzing the time
evolution of the trajectory for each process more precisely. As
discussed in reference \cite{ari04}, the difference between the QF
and DQF process is the mass asymmetry of the fission fragments.
However, in the previous work, the origin of the mass asymmetry of
the fission fragments in the both processes never have been
discussed. It should be discussed what is a necessary condition
for the trajectory to take the fusion path.

We aim to clarify the following points: (i)~the mechanism that
controls each path, (ii)~the conditions governing the different
paths, (iii)~the conditions necessary to follow the fusion path.
The last item is strongly related to attempts to synthesize new
superheavy elements.

In section~2, we explain the framework of our study and the model
used. We discuss the fusion mechanism in the reaction
$^{48}$Ca+$^{208}$Pb in section~3. In section~4, we analyze the
trajectory in reaction $^{48}$Ca+$^{244}$Pu precisely. We clarify
the mechanism for each process. The origins of the FF, DQF and QF
processes are discussed in section~5. We also analyze the
processes in terms of  the nuclear shape for each path in
section~6. In section~7, the dependence of the trajectory on
incident energy is investigated. We present a summary and further
discussion in section~8.

\section{Model}

Using the same procedure as described in reference \cite{ari04} to
investigate the fusion-fission process dynamically, we use the
fluctuation-dissipation model and employ the Langevin equation. We
adopt the three-dimensional nuclear deformation space given by
two-center parameterization \cite{maru72,sato78}. The three
collective parameters involved in the Langevin equation are as
follows: $z_{0}$ (distance between two potential centers),
$\delta$ (deformation of fragments) and $\alpha$ (mass asymmetry
of the colliding nuclei); $\alpha=(A_{1}-A_{2})/(A_{1}+A_{2})$,
where $A_{1}$ and $A_{2}$ denote the mass numbers of the target
and the projectile, respectively. In two-center shell model, it is
difficult to express the realistic nuclear shape with the
approximately $|\alpha| > 0.5$. In sections~3 and 6 below, this
difficulty is mentioned.

The parameter $\delta$ is defined as $\delta=3(a-b)/(2a+b)$, where
$a$ and $b$ denote the half length of the axes of ellipse in the
$z$ and $\rho$ directions, respectively as expressed in Fig.~1 in
reference \cite{maru72}. We assume that each fragment has the same
deformations as the first approximation. The neck parameter
$\epsilon$ is the ratio of the smoothed potential height to the
original one where two harmonic oscillator potentials cross each
other. It is defined in the same manner as reference
\cite{maru72}. In the present calculation, $\epsilon$ is fixed to
be 1.0, so as to retain the contact-like configuration more
realistically for two-nucleus collision.

The multidimensional Langevin equation is given as
\begin{eqnarray}
\frac{dq_{i}}{dt}&=&\left(m^{-1}\right)_{ij}p_{j},\nonumber\\
\frac{dp_{i}}{dt}&=&-\frac{\partial V}{dq_{i}}
                 -\frac{1}{2}\frac{\partial}{\partial q_{i}}
                   \left(m^{-1}\right)_{jk}p_{j}p_{k}
                  -\gamma_{ij}\left(m^{-1}\right)_{jk}p_{k}
                  +g_{ij}R_{j}(t),
\end{eqnarray}
where a summation over repeated indices is assumed. $q_{i}$
denotes the deformation coordinate specified by $z_{0}$, $\delta$
and $\alpha$. $p_{i}$ is the conjugate momentum of $q_{i}$. $V$ is
the potential energy, and $m_{ij}$ and $\gamma_{ij}$ are the
shape-dependent collective inertia parameter and dissipation
tensor, respectively. A hydrodynamical inertia tensor is adopted
in the Werner-Wheeler approximation for the velocity field, and
the wall-and-window one-body dissipation is adopted for the
dissipation tensor \cite{bloc78,nix84,feld87}. The normalized
random force $R_{i}(t)$ is assumed to be  white noise, {\it i.e.},
$\langle R_{i}(t) \rangle$=0 and $\langle R_{i}(t_{1})R_{j}(t_{2})
\rangle = 2 \delta_{ij}\delta(t_{1}-t_{2})$. The strength of
random force $g_{ij}$ is given by $\gamma_{ij}T=\sum_{k}
g_{ij}g_{jk}$, where $T$ is the temperature of the compound
nucleus calculated from the intrinsic energy of the composite
system as $E_{int}=aT^2$, with $a$ denoting the level density
parameter. The temperature-dependent potential energy is defined
as

\begin{equation}
V(q,l,T)=V_{DM}(q)+\frac{\hbar^{2}l(l+1)}{2I(q)}+V_{shell}(q)\Phi
(T),
 \label{vt1}
\end{equation}
\begin{equation}
V_{DM}(q)=E_{S}(q)+E_{C}(q),
\end{equation}
where $I(q)$ is the moment of inertia of a rigid body at
deformation $q$,  $V_{shell}$ is the shell correction energy at
$T=0$, and $V_{DM}$ is the potential energy of the finite-range
liquid drop model. $E_{S}$ and $E_{C}$ denote a generalized
surface energy \cite{krap79} and Coulomb energy, respectively. The
centrifugal energy arising from the angular momentum $l$ of the
rigid body is also considered. The temperature-dependent factor
$\Phi$ is parameterized as $\Phi=$exp$\{-aT^{2}/E_{d}\}$ following
the work of Ignatyuk et al. \cite{ign75}. The shell dumping energy
$E_{d}$ is chosen to be 20 MeV. The intrinsic energy of the
composite system $E_{int}$ is calculated for each trajectory as

\begin{equation}
E_{int}=E^{*}-\frac{1}{2}\left(m^{-1}\right)_{ij}p_{i}p_{j}-V(q,l,T),
\end{equation}

where $E^{*}$ denotes the excitation energy of the compound
nucleus, and is given by $E^{*}=E_{cm}-Q$ with $Q$ and $E_{cm}$
denoting the $Q-$value of the reaction and the incident energy in
the center-of-mass frame, respectively.

Also, we take into account the neutron emission in the Langevin
calculation during the fusion-fission process, in the same manner
as reference \cite{frob96}.


\section{Fusion mechanism in the reaction $^{48}$Ca+$^{208}$Pb}

In our first attempt, we investigate the reaction
$^{48}$Ca+$^{208}$Pb, in which the FF process is dominant, as
shown in reference \cite{ari04}. In this reference, we have
already compared the calculation with the experimental data of the
mass distribution of fission fragments and fusion-fission cross
section, and the calculation shows a good agreement with the
experimental data. We took into account the contribution of
various angular momentum in reference \cite{ari04}. Here, in order
to investigate the behavior of the trajectory in the dynamical
process, we mainly discuss for zero angular momentum case.

Figure~1 shows the potential energy surface of the liquid drop
model with the shell correction energy for $^{256}$No on the
$z-\alpha$ $(\delta=0$) plane, for the case of $l=0$. This
potential energy surface is calculated using the two-center shell
model code \cite{suek74,iwam76}. The contour lines of the
potential energy surface are drawn in steps of 2~MeV. Like the
deformation parameters of nuclear shape described in reference
\cite{ari04}, $z$ is defined as $z=z_{0}/(R_{CN}B)$, where
$R_{CN}$ denotes the radius of the spherical compound nucleus.
Parameter $B$ is defined as $B=(3+\delta)/(3-2\delta)$.


\begin{figure}
\centerline{
\includegraphics[height=.41\textheight]{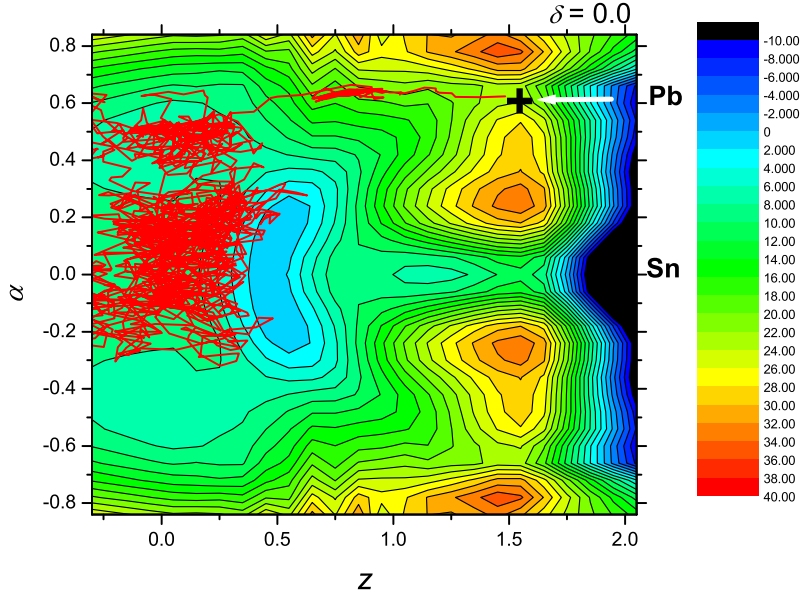}}
  \caption{Samples of the trajectory of the fusion
process projected onto $z-\alpha$ $(\delta=0)$ plane at $E^{*}=50$
MeV in reaction $^{48}$Ca+$^{208}$Pb. The potential energy surface
is presented by a liquid drop model with shell correction energy
in the nuclear deformation space for $^{256}102$. The white arrow
shows the injection point of this reaction. Symbols are
 given in the text.}
\end{figure}


\begin{figure}
\centerline{
\includegraphics[height=.75\textheight]{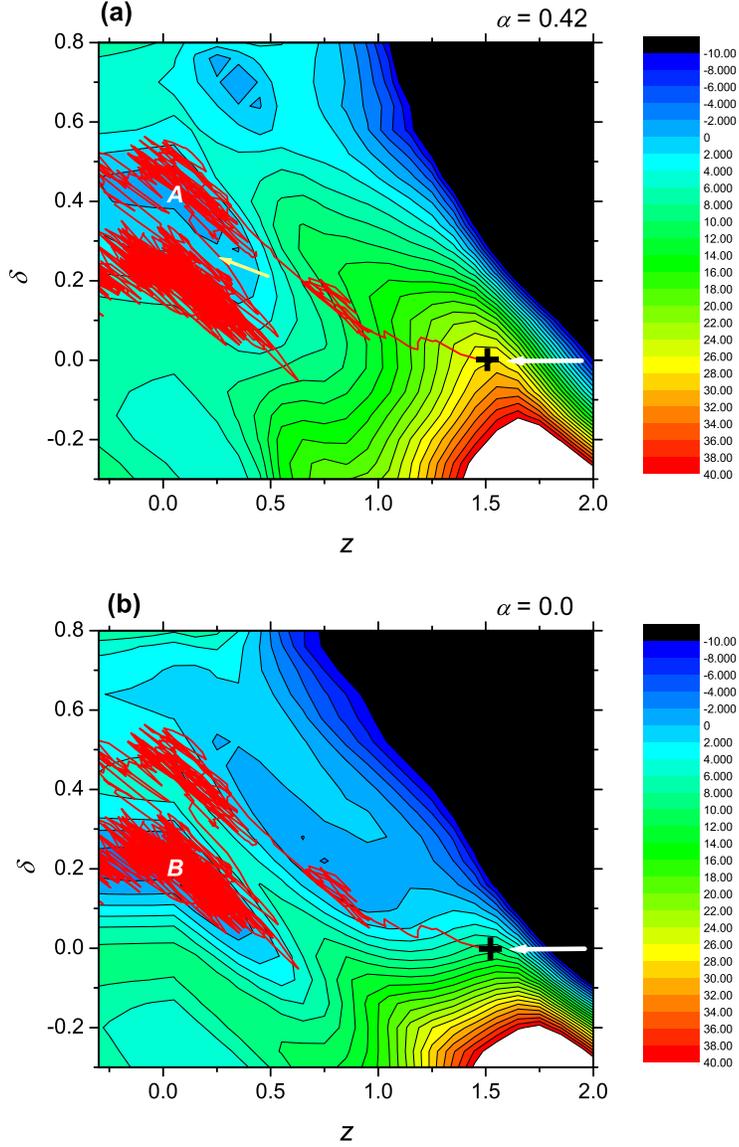}}
  \caption{Sample of the trajectory projected onto $z-\delta$
  plane in reaction $^{48}$Ca+$^{208}$Pb at $E^{*}=50$ MeV.
 (a) $\alpha=0.42$. (b) $\alpha=0.0$. The subpocket $(A)$ appearing at
 a large value of $\alpha$ disappears gradually with decreasing
$\alpha$, and moves into the main pocket $(B)$. The fusion
trajectory is trapped at $(A)$ and $(B)$ to form a compound
nucleus. Symbols are given in the text.}
\end{figure}



In Fig.~1, the position at $z=\alpha=0$ corresponds to a spherical
compound nucleus. The injection point of this system is indicated
by the arrow. The (+) denotes the point of contact in the system.
We start to calculate the three-dimensional Langevin equation at
the point of contact located at $z=1.56, \delta=0.0, \alpha=0.62$.
All trajectories start at this point with the momentum in the
initial channel.

Due to the shell structure of the nucleus, we can see a valley in
the potential energy surface near Pb and Sn fragments, as
indicated in Fig.~1. The valley leads to the spherical region from
the point of contact. Therefore, as shown in Fig.~1, most
trajectories travel down along the valley and reach the spherical
region. The red line in Fig.~1 denotes a sample trajectory of the
fusion process at the incident energy corresponding to the
excitation energy of the compound nucleus $E^{*}=50$ MeV, which is
projected onto the $z-\alpha$ ($\delta=0$) plane. The trajectories
of the FF process make up 96.2 \% of all trajectories.

However, when we plot the trajectory on the $z-\delta$ plane, the
process seems not so simple. Figure~2 shows the potential energy
surface on the $z-\delta$ plane. We can see the steep descending
slope in the direction $+\delta$ near the point of contact in the
system. The same trajectory shown in Fig.~1 is plotted on the
different two planes. The trajectory flows down along the slope in
the $+\delta$ direction.





\begin{figure}
\centerline{
\includegraphics[height=.79\textheight]{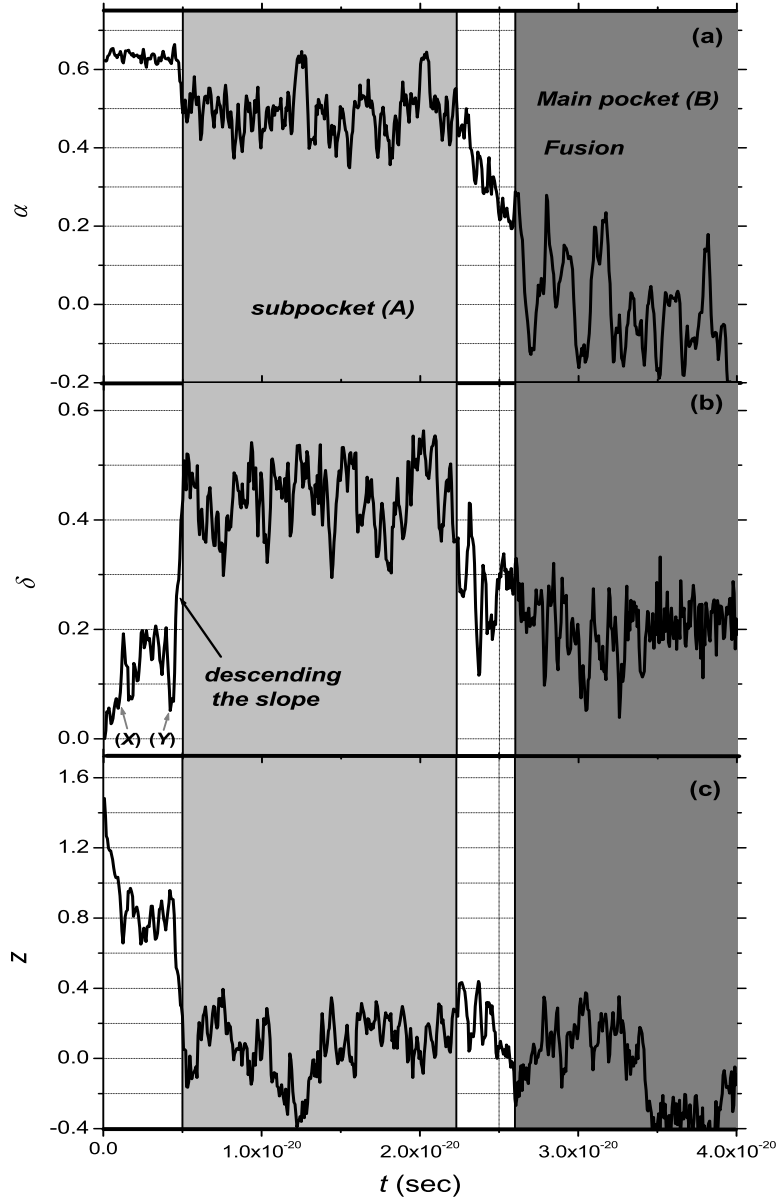}}
  \caption{The time evolution of each parameter for the sample
trajectory shown in Fig.~1 in
  reaction $^{48}$Ca+$^{208}$Pb at
  $E^{*}=50$ MeV. (a) Parameter $\alpha$. (b) Parameter $\delta$.
  (c) Parameter $z$. Symbols are given in the text.}
\end{figure}

Figures~2(a) and (b) show the potential energy surface on the
$z-\delta$ plane at $\alpha=0.42$ and $\alpha=0.0$, respectively.
In Fig.~2(a), this trajectory is trapped in the pocket located at
$z\sim 0.2, \delta \sim 0.4, \alpha \sim 0.4$ (indicated by $A$).
This pocket is produced by the shell effect of the deformed
nucleus. Here, we call this pocket the subpocket. Then, the
trajectory moves to another pocket located near the spherical
region (indicated by $B$) in Fig.~2(b), and is trapped there. We
call this pocket the main pocket. It is very interesting that, in
this reaction, the fusion process proceeds in two steps, that is,
the trajectory is trapped in the subpocket~$(A)$ first and then
moves to the main pocket~$(B)$ associated with the process of
relaxation of mass asymmetry. In Fig.~2(a), the yellow arrow
denotes the transition process from the subpocket with the
deformed shape to the main pocket of the spherical shape.

As discussed in reference \cite{ari04}, the trajectory is
prevented from going to the fusion area by the steep slope in the
$+\delta$ direction. However, in this case, there is a pocket in
the large $+\delta$ area, as shown in Fig.~2(a). Trajectories
flowing to the $+\delta$ direction are trapped in the subpocket
for a moment and are blocked from going to the fission area. We
prepare 1,000 trajectories in this calculation, and find that all
trajectories of the FF process are this type of trajectory.


\begin{figure}
\centerline{
\includegraphics[height=.75\textheight]{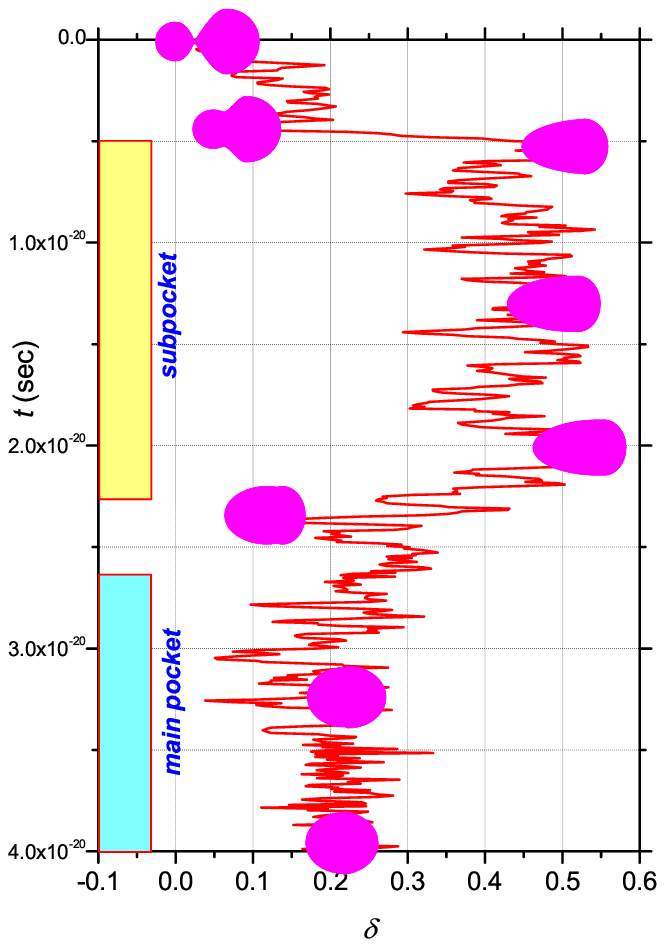}}
  \caption{The time evolution of parameter $\delta$ and
the nuclear shape at each deformation point for the sample
trajectory shown in Fig.~1 in reaction $^{48}$Ca+$^{208}$Pb at
$E^{*}=50$ MeV.}
\end{figure}

As mentioned above, we assume that kinetic energy does not
dissipate during the approaching process. At the point of contact,
we start the Langevin calculation.  Due to the strong friction,
all kinetic energy dissipates in a very short time scale, that is,
by $t=3.0 \times 10^{-22}$ sec. The time evolutions of parameters
$\alpha, \delta$ and $z$ for the sample trajectory in Fig.~1 are
shown in Figs.~3(a), (b) and (c), respectively. The periods when
the trajectory is trapped in the subpocket and the main pocket are
presented by the light gray and gray areas, respectively. It is
clearly shown that the fusion trajectory is trapped by the
subpocket (staying for a time interval of $1.7 \times 10^{-20}$
sec) and then moves to the main pocket. In Fig.~3(c), until all
the kinetic energy dissipates, parameter $z$ moves with very high
speed in the $-z$ direction, because it has initial momentum in
this direction. After it reaches $z \sim 1.2$, the trajectory
begins to descend along the steep slope in the $+\delta$ direction
(indicated by $(X)$ in Fig.~3(b)). After $t=4.5 \times 10^{-21}$
sec, it quickly descends the slope at very high speed (indicated
by $(Y)$ in Fig.~3(b)). At $t=5.0\times 10^{-21}$ sec, the
trajectory reaches the subpocket and is trapped. During the stay
in the pocket, the mass-asymmetry parameter $\alpha$ fluctuates
around $\alpha\sim 0.5$, and it is relaxed quickly when the
trajectory begins to move to the main pocket. At the same time,
the value of $\delta$ decreases, and due to the relaxation of
$\alpha$, the main pocket becomes larger and deeper. Therefore,
the trajectory is trapped in it for a long time. At $t=2.6 \times
10^{-20}$ sec, it enters the main pocket, which is regarded as the
accomplishment of fusion.

Figure~4 shows the time evolution of parameter $\delta$ and the
nuclear shape at each specific point for the sample trajectory
shown in Fig.~1. These shapes are presented using the two-center
shell model parameterization. As mentioned in the previous
section, due to the difficulty with the parametrization with the
approximately $|\alpha| > 0.5$, the nuclear shape is not exactly
spherical-spherical at the point of the contact ($\alpha=0.62$).

At $t=5.0 \times 10^{-21}$ sec, by increasing the value of
$\delta$ and decreasing $z$ quickly, the neck of the nuclear shape
disappears. We can see that the shape is stabilized by the shell
structure of the deformed nucleus in this case. Then, at $t \sim
2.25 \times 10^{-20}$, the trajectory moves to the main pocket,
which fusion time corresponds to that calculated by Blocki et. al.
\cite{bloc86}. The shape in the pocket is more compact, and this
nucleus is considered to be a compound nucleus. Thus, in this
reaction, the parameter $\delta$ plays a very important role in
the fusion reaction.

In this analysis, we reveal the fusion mechanism in this reaction.
Such a mechanism cannot be revealed by analyzing only the process
on the $z-\alpha$ plane.



\begin{figure}
\centerline{
\includegraphics[height=.75\textheight]{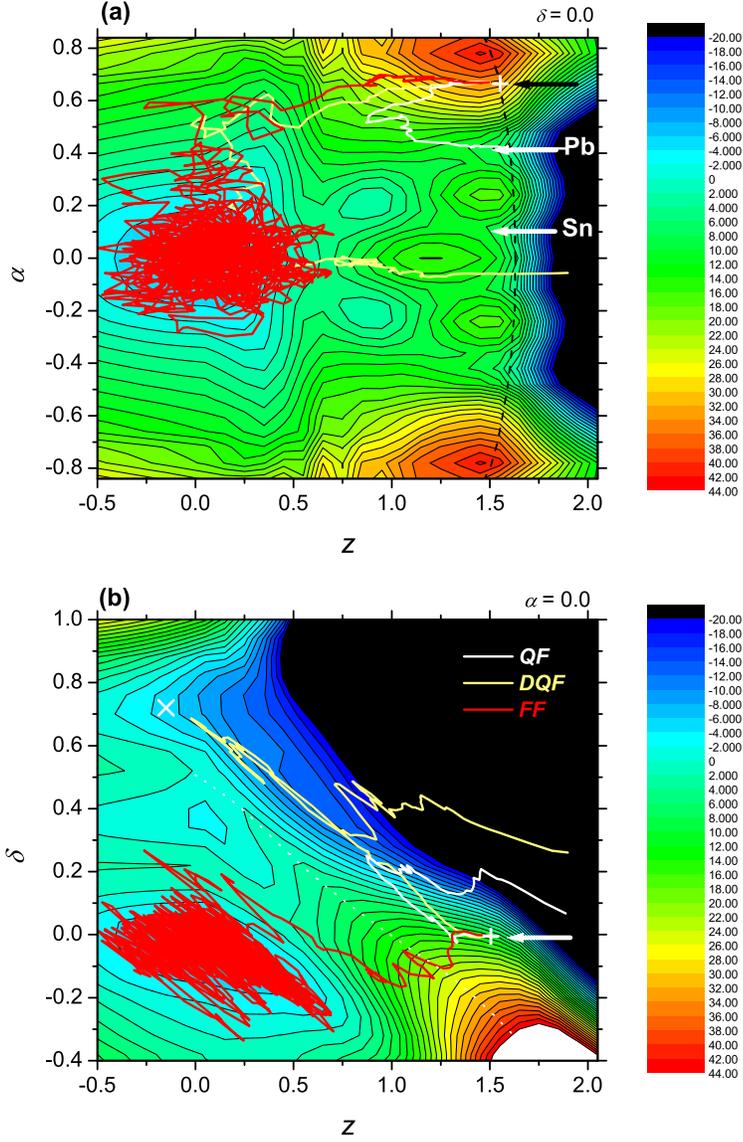}}
  \caption{Samples of the trajectory projected onto  $z-\alpha$
  $(\delta=0)$ plane (a) and $z-\delta$
  $(\alpha=0)$ (b) at $E^{*}=50$ MeV in reaction $^{48}$Ca+$^{244}$Pu.
The trajectories of the QF, DQF and FF processes are denoted by
the white, yellow and red lines, respectively. The potential
energy surface is presented by the liquid drop model with shell
correction energy in the nuclear deformation space for
$^{292}114$. In (a), the line of the point of contact is denoted
by the black broken line. In (b), the white broken line denotes
the ridge line. Symbols are given in the text.}
\end{figure}

\section{Behavior of each path in the reaction $^{48}$Ca+$^{244}$Pu}

In our previous work \cite{ari04}, we have presented that the good
agreements of the calculation with the experimental cross section
derived by the mass symmetric fission events in the reaction
$^{48}$Ca+$^{244}$Pu. Also we classified the dynamical process on
the basis of the features of the trajectory corresponding to the
characteristic physical phenomena. They are the QF, DQF and FF
processes. Within the model, here we analyze these different
trajectories more precisely and reveal the mechanism by which the
trajectory chooses its characteristic path. We investigate the
time evolution for each trajectory and try to find the condition
under which the trajectory takes the fusion path. In particular,
the required conditions for inducing the fusion path are one of
the important factors in the synthesis of superheavy elements. In
order to discuss the dynamical process, we mainly focus on zero
angular momentum case. The main mechanism of fusion-fission
process can be explained by the trajectory's behavior on the
potential landscape. For each angular momentum, the potential
landscape changes but the mechanism of the process essentially
does not change.

Figure~5 shows the potential energy surface of the liquid drop
model with shell correction energy for $^{292}114$ on the
$z-\alpha$ $(\delta=0)$ plane (a) and the $z-\delta$ $(\alpha=0)$
plane (b) in the case of $l=0$. The white, yellow and red lines
denote sample trajectories of the QF, DQF and FF processes,
respectively, in the reaction $^{48}$Ca+$^{244}$Pu at the incident
energy corresponding to the excitation energy of the compound
nucleus $E^{*}=50$ MeV. These trajectories are projected onto the
$z-\alpha$ ($\delta=0$) plane and $z-\delta$ ($\alpha=0$) plane,
as shown in Fig.~5(a) and (b), respectively.

In this case, most of the trajectories enter into the fission area
by passing along the QF path, thus, the QF process is dominant.
One of the fission fragments is distributed around Pb, which is
marked in Fig.~5(a) with the white arrow. In this calculation, we
assume that the shapes of both the target and the projectile are
spherical at the point of contact of the colliding system, even
though $^{244}$Pu is a deformed nucleus. We define the fusion box
as the inside of the fission saddle point on each axis, $\{z <
0.6, \delta < 0.2, |\alpha | < 0.25\}$ \cite{ari04}.
The injection point of this system is indicated by the black arrow
in Fig.~5(a) and the white arrow in Fig.~5(b). We start to
calculate the three-dimensional Langevin equation at the point of
contact marked by (+), which is located at $z=1.54, \delta=0.0,
\alpha=0.67$.


Figure 6 shows the time dependence of each deformation parameter
for the sample trajectories shown in Fig.~5.  The time evolutions
of parameters $\alpha,\delta$ and $z$ are shown in (a),(b) and
(c), respectively. In each figure, the green, blue and red lines
denote the QF, DQF and FF processes, respectively.

By analyzing the time evolution of each parameter and the
relationships among them, we can see the characteristic behavior
of each path. As explained in the following subsection, we found
that the most important stage in determining the process is the
stage from the point of contact to just before the ridge line on
the $z-\delta$ plane indicated by the white broken line in
Fig.~5(b). Here, we call this stage the critical stage. First, we
discuss the feature of each path by analyzing the trajectories in
Fig.~5, and then we discuss this critical stage in the next
section.

\begin{figure}
\centerline{
  \includegraphics[height=.75\textheight]{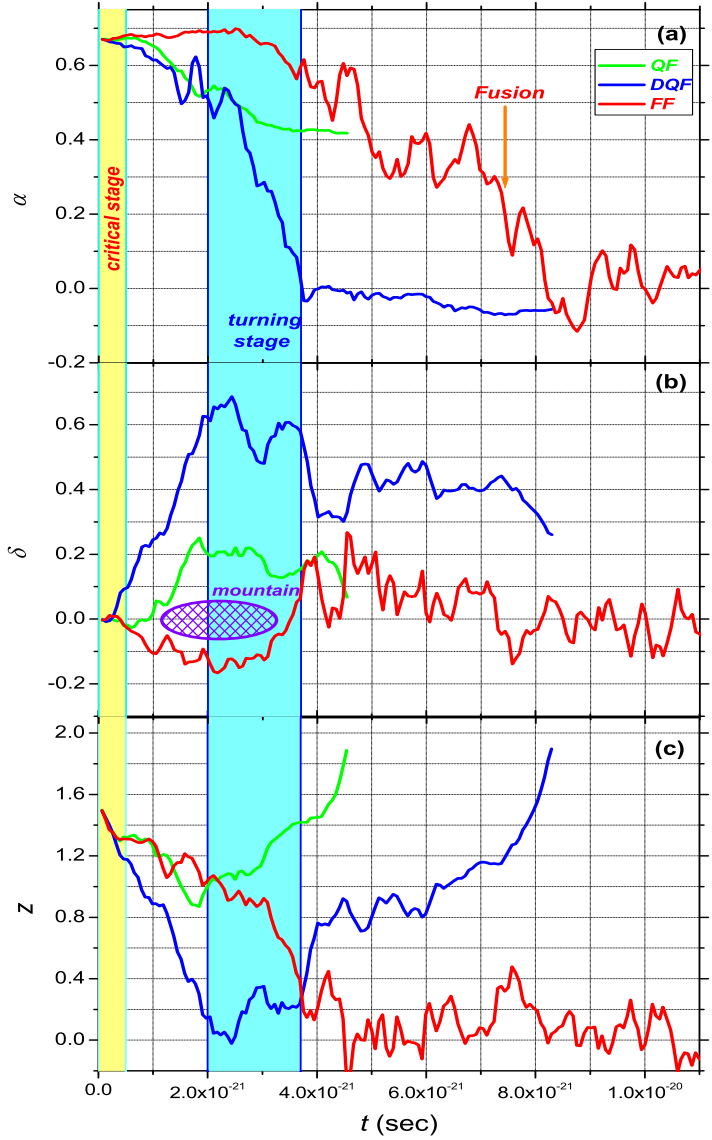}}
  \caption{Time dependence of each deformation parameter for the QF, DQF and FF
  paths shown in Fig.~5. They are denoted by the green, blue and red
lines, respectively. The reaction is $^{48}$Ca+$^{244}$Pu at
$E^{*}=50$ MeV. }
\end{figure}


\begin{figure}
\centerline{
\includegraphics[height=.37\textheight]{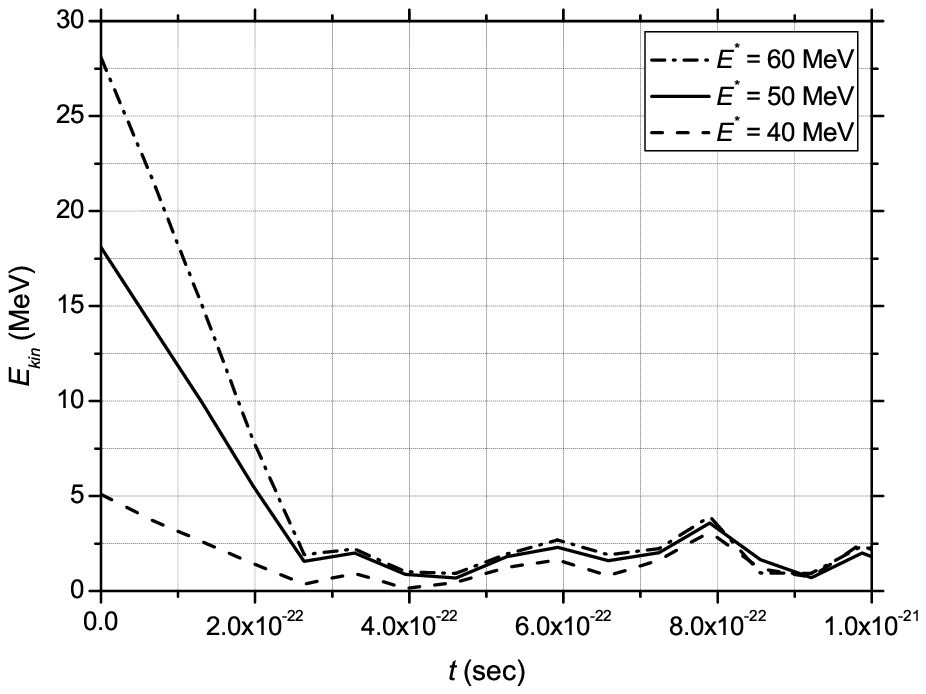}}
  \caption{Samples of the time evolutions of the kinetic energy of
  relative motion in reaction $^{48}$Ca+$^{244}$Pu at
  $E^{*}=40, 50$ and 60 MeV, which are denoted by the dashed,
  solid and dashed-dotted lines, respectively.}
\end{figure}


\subsection{Fusion-fission path}

As discussed in the previous section, we assume that the kinetic
energy does not dissipate during the approaching process. The
initial velocity is directed in only the $-z$ direction, and the
two components of the initial velocities along the $\delta$ and
$\alpha$ directions are assumed to be zero. Hence, first, all
trajectories move in the $-z$ direction and overcome the ridge
near $z \sim 1.45 $ on the $z-\alpha$ plane, as shown in
Fig.~5(a).

After overcoming the ridge, at $t=0.5 \times 10^{-21}$ sec, the
trajectories approach to the ridge on the $z-\delta$ plane denoted
by the white broken line in Fig.~5(b). The momentum of this
trajectory is decreased already due to strong friction. Then, by
random force perpendicular to the ridge line, the trajectory
overcomes the ridge.
Overcoming the ridge is the key to
whether the trajectory follows the fusion process.

Once the trajectory overcomes the ridge on the $z-\delta$ plane,
it descends along the valley to the spherical area.
This valley is located in the $-\delta$ region, so that the
trajectory is blocked by a mountain from going to the $+\delta$
direction which leads to the fission area, as shown in Fig.~6(b).

The trajectory goes along the valley until $t=3.7 \times 10^{-21}$
sec, and reaches $z=0.4$ which is the border of the fusion box on
the $z$ coordinate. During this process, the mass-asymmetry
parameter $\alpha$ does not change very much, as shown in
Figs.~6(a). Parameter $\delta$ also does not change very much
during this time, because the trajectory descends along the valley
which is located in the $-\delta$ region.

After the trajectory enters the $z< 0.4$ area, the mass asymmetry
$\alpha$ begins to relax and the trajectory enters the fusion box
at $t=7.5 \times 10^{-21}$ sec, indicated by the arrow in
Fig.~6(a). Here, the important point is that the mass asymmetry
$\alpha$ relaxes after $z$ becomes sufficiently small. The
trajectory is trapped in the pocket near the spherical region.

\subsection{Deep quasi-fission path}

The trajectory of the DQF process, after starting at the point of
contact, also first overcomes the ridge on the $z-\alpha$ plane.
Immediately afterwards, the trajectory begins to avoids the ridge
on the $z-\delta$ plane. In this case, the momentum in the $-z$
direction of the trajectory is not lost so much and the trajectory
slips through the lower foot of the ridge on the $z-\delta$ plane
at a high speed and descends the steep slope in the $+\delta$
direction. Due to the momentum being in the $-z$ direction,
parameter $z$ also decreases quickly. This high-speed process
proceeds until $t=2.0 \times 10^{-21}$ sec, as shown in Figs.~6(b)
and (c). The value of $z$ becomes almost zero during this time and
the deformation of fragments occurs up to $\delta \sim 0.6$.

Finally the trajectory reaches the area where the potential energy
surface becomes gentle, as indicated by $(\times)$ in Fig.~5(b).
In this area, the trajectory speed in the $z$ and $\delta$
directions decreases and becomes almost zero on the $z-\delta$
plane. The movement of the trajectory undergoes a change to the
$+z$ direction. During the time taken for this turning process,
the mass asymmetry is drastically relaxed, as shown in Fig.~6(a).
The time taken for this process is $2.0 \sim 3.7 \times 10^{-21} $
sec. We call this period the turning stage which is shown in
Fig.~6. On the other hand, neither parameter $\delta$ nor $z$
changes very much during this time. We can say that the turning
point of the DQF path exists at $z\sim 0.0, \delta=0.5\sim 0.6,
\alpha= 0.4\sim 0.5$.

After the trajectory completes its change of direction to $+z$, it
travels down along the steep slope in the $+z$ direction. In this
fission process, the trajectory goes out of the fission area with
a large value of $\delta$.

\subsection{Quasi-fission path}

The trajectory of the QF path is the same as that of the FF path,
up to $t=0.5 \times 10^{-21}$ sec. That is to say, the trajectory
travels in the $-z$ direction with substantial momentum and
overcomes the ridge on the $z-\alpha$ plane. Then, the trajectory
reaches the point near the ridge on the $z-\delta$ plane, which is
the same point as that in the FF path. Due to the strong friction,
the momentum of this trajectory is lost during this process. As
shown in Fig.~6(c), the trajectory almost stops at this moment.
The decrease of $z$ saturates.

Subsequently, random force controls the future behavior of the
trajectory. Different from the FF case, the trajectory is
proceeded by random forces whose direction has a low angle with
respect to the ridge line or are orienting to the opposite side of
the ridge. It does not overcome the ridge line, and falters and
descends quickly along the steep slope in the $+\delta$ direction.
During this short process, there is not sufficient time for the
relaxation of the mass-asymmetry parameter $\alpha$. During the
fission process, $\delta$ does not change greatly and retains the
value of $\delta \sim 0.2$.

\section{Key factor in each process}

In this section, we discuss the mechanism of the path separation.
We investigate the origin of the QF, DQF and FF paths. By
analyzing the trajectories in the previous sections, all processes
are found to easily overcome the ridge on the $z-\alpha$ plane,
which we call the first ridge. We found that the most important
stage (critical stage) in deciding the branch of the path is the
process from the point of contact to the next ridge on the
$z-\delta$ plane, which we call the second ridge. This critical
stage is shown in Fig.~6.


\begin{figure}
\centerline{
\includegraphics[height=.51\textheight]{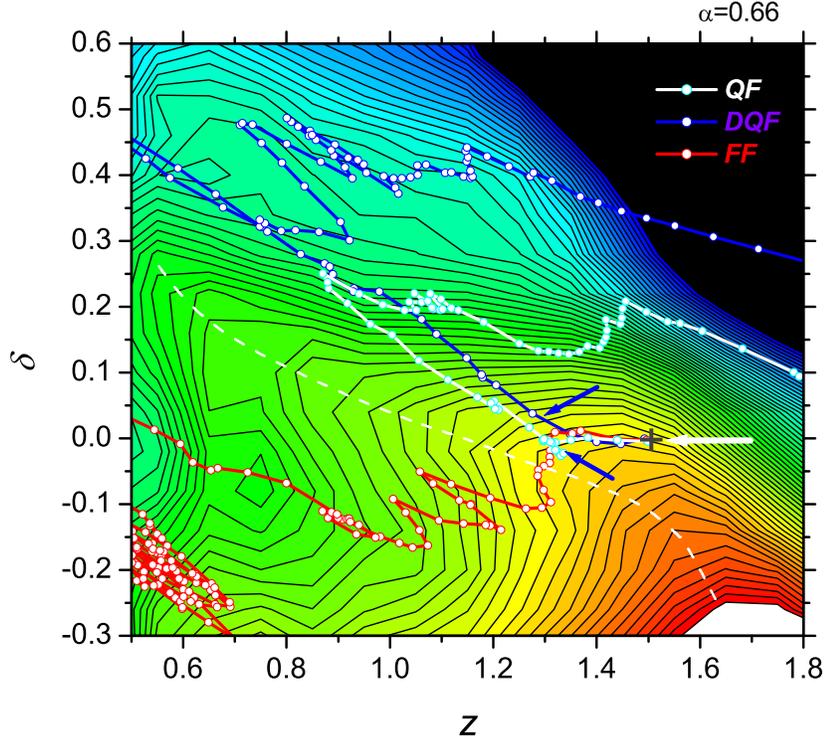}}
  \caption{Samples of the trajectory near the point of contact,
   projected onto $z-\delta$ $(\alpha=0.66)$ plane at $E^{*}=50$ MeV.
  The QF, DQF and FF processes are denoted by the
  white, blue and red lines, respectively. The white dashed line denotes
  the second ridge line. The blue arrows show the points of $t_{diss}$.
  The dots correspond to the time steps of $t=6.58\times 10^{-23}$ sec.}
\end{figure}

Figure~7 shows the samples of the time evolution of the kinetic
energy of the relative motion for a trajectory in this reaction.
The solid line denotes the case of $E^{*}=50$ MeV. The kinetic
energy dissipates at $t=3.0 \times 10^{-22}$ sec. Up to this time,
the trajectory advances in the $-z$ direction and overcomes the
first ridge on the $z-\alpha$ plane at a location very near the
point of contact.

After overcoming this barrier, the next process is very important
in deciding the bifurcation. The trajectories of the FF and QF
paths advance toward the point near the second ridge line on the
$z-\delta$ plane. Figure~8 shows each trajectory near the point of
contact. These trajectories are plotted on the $z-\delta$ plane
$(\alpha=0.66)$. The QF, DQF and FF processes are denoted by the
white, blue and red lines, respectively. The open circles on each
trajectory denote the time step. The interval of two circles mark
corresponds to the time step of $6.58 \times 10^{-23}$ sec. The
white dashed line denotes the second ridge line. In Fig.~8, the
positions at which all the kinetic energy dissipates are marked by
the blue arrows.

The trajectories of the FF and QF paths lose all the kinetic
energy near the second ridge on the $z-\delta$ plane. At these
points, random force determines whether the trajectory overcomes
the the second ridge. By random force, if trajectory overcomes the
second ridge, it is the FF process. The trajectory of the DQF path
begins to avoid the lower foot of the second ridge on the
$z-\delta$ plane just after overcoming the first ridge with
momentum in the $-z$ direction.

The process is mainly decided in the critical stage. Therefore, it
is worth to remark that the effective fusion barrier is considered
to be the second ridge on the $z-\delta$ plane and not the first
ridge near the point of contact. Even if the trajectory overcomes
the first ridge, it does not mean that it enters the fusion area.
We can say that if the trajectory overcomes the second ridge line,
it will enter the fusion region. This is the necessary condition
for the trajectory to enter the fusion area in the superheavy-mass
region.

Actually, in the three-dimensional coordinate space, the process
of the trajectory has already been decided near the point of
contact. In the early stages, the trajectory is separated into the
FF, DQF and QF processes. The first ridge line is related to the
deep inelastic collision (DIC) process. When the trajectory cannot
overcome the first ridge line, it is recognized as a DIC process.

\section{Time evolution of the nuclear shape in each process}

In this section, we  investigate the different paths from the
viewpoint of nuclear shape. We compare the time evolutions of the
nuclear shapes in different paths, keeping in mind the discussion
in the previous section.


\begin{figure}
\centerline{
\includegraphics[height=.75\textheight]{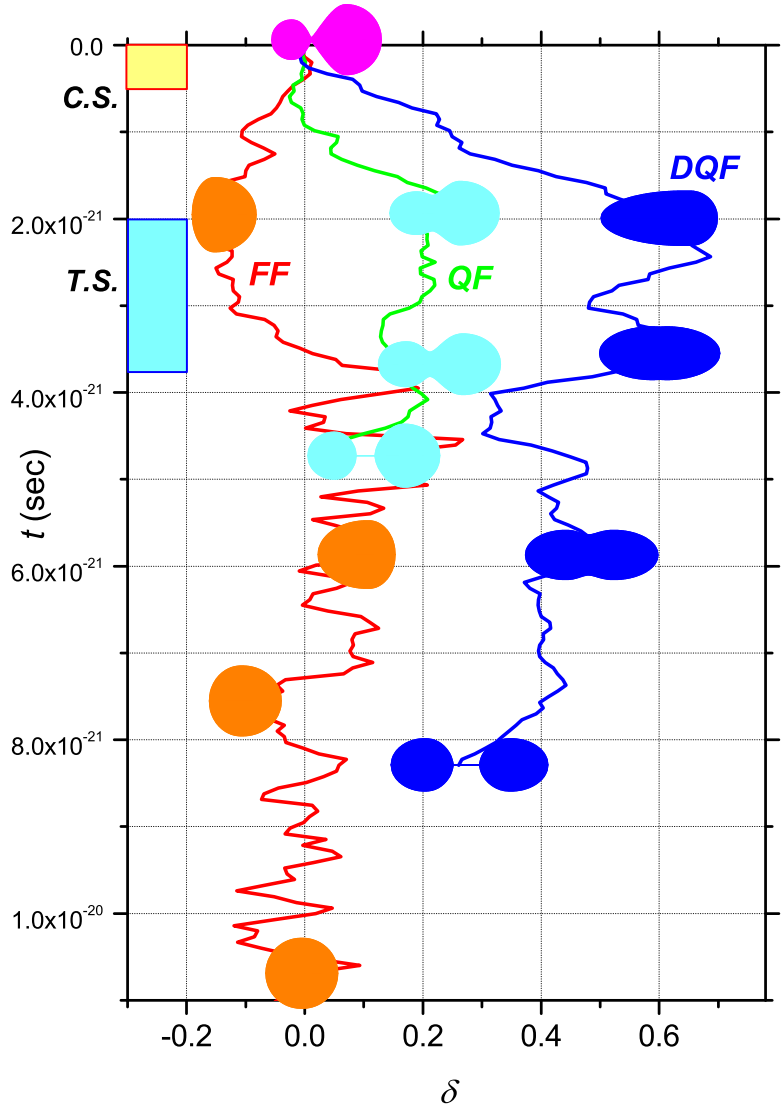}}
  \caption{The time evolutions of parameter $\delta$ and
the nuclear shapes at several points for the trajectories shown in
Fig.~5, in reaction $^{48}$Ca+$^{244}$Pu at $E^{*}=50$ MeV. The
process is indicated by $QF, DQF$ and $FF$. The blocked areas
indicated by $C.S.$ and $T.S.$ are the critical stage and the
turning stage, respectively, corresponding to those in Fig.~6.}
\end{figure}


Figure~9 shows the time evolution of parameter $\delta$ of each
path with the nuclear shapes at several deformation points for the
same trajectories in Fig.~5. As mentioned above, because of the
difficulty with the parametrization with the approximately
$|\alpha| > 0.5$, the nuclear shape is not realistic near the
point of the contact ($\alpha=0.67$). However, in the sets of our
parameters, the potential energy at the point of the contact is
consistent with that calculated by the Bass model \cite{bass741}.
The red, blue and green lines denote the FF, DQF and QF processes,
respectively. The blocked areas indicated by $C.S.$ and $T.S.$ are
the critical stage and the turning stage, respectively,
corresponding to those in Fig.~6. After passing the critical
stage, each process takes a different value of $\delta$. Only the
trajectory of the FF process takes a negative value of $\delta$,
and it is very important for the fusion process to maintain the
oblate fragment deformation until the relaxation of $\alpha$, as
mentioned in the previous section. In the FF process, up to
$t=4.0\times 10^{-21}$ sec, parameter $z$ approaches $z=0$, so
that the shape at this point is very compact even if the mass
asymmetry $\alpha$ is large. This is one of the restrictions of
the two-center shell model parameterization. Subsequently, in the
FF process, the nuclear shape fluctuates around a sphere due to
the thermal fluctuation.

In the QF process, at $t=2.0\times 10^{-21}$ sec, parameter $z$ is
about $z=1.0$, as shown in Fig.~6, and this value is similar that
in the FF process. Since the value of the parameter $\delta \sim
0.2$, the nuclear shape has a neck, as shown at this point in
Fig.~9. This neck is considered to contribute to faster fission
compared with the DQF process, which does not have a neck at this
time. Therefore, the trajectory progresses to fission keeping its
mass asymmetry.

Regarding the DQF process, at $2.0\times 10^{-21}$ sec, due to the
large value of $\delta$ and the small value of $z$, the neck
disappears and the nuclear shape resembles a mononucleus. From
this shape, it appears to be easy to change to the mass-symmetric
shape. During the turning stage, the mass asymmetry of this
nucleus is relaxed completely. The time interval of this process
is about $1.8\times 10^{-21}$ sec. Due to large deformation, even
if the mass asymmetry is relaxed, the trajectory cannot reach the
fusion area. In this case, the mass-symmetric fission occurs
without forming a compound nucleus.



\section{Dependence of incident energy}

In this section, we discuss how the contribution from each process
changes depending on the incident energy in the reaction
$^{48}$Ca+$^{244}$Pu.


\begin{figure}
\centerline{
\includegraphics[height=.840\textheight]{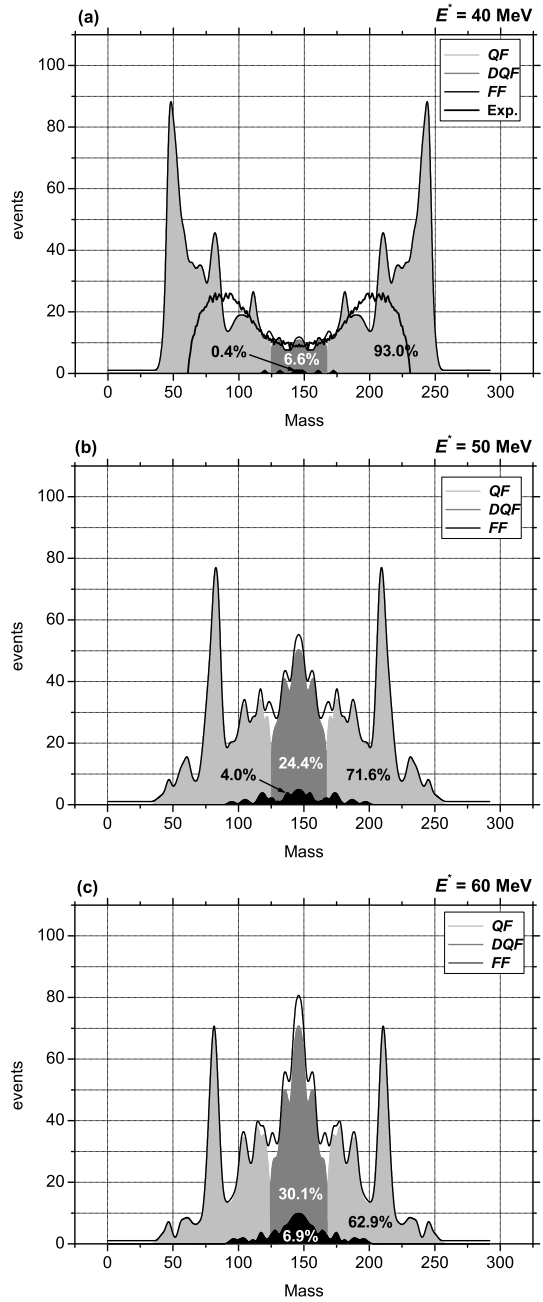}}
  \caption{The mass distributions of fission fragments in reaction
  $^{48}$Ca+$^{244}$Pu, which are distinctive for each process.
   Figures denotes the probability of each
process. (a), (b) and (c) are shown at $E^{*}=40,50$ and 60 MeV,
respectively.}
\end{figure}


Figure~10 shows the mass distributions of fission fragments which
are distinctive for each process. The fission fragments from the
QF, DQF and FF processes are presented by the light gray, gray and
black shadings, respectively. In our calculation, the events of QF
in Fig.~10 include some events of the DIC process. The thin black
line denotes the total process. Figures denote the probability of
each process, which are given in \%. Three panels (a),(b) and (c)
show the distribution of fission fragments for the incident energy
corresponding to the excitation energy of the compound nucleus
$E^{*}=40,50$ and 60 MeV, respectively.

At $E^{*}=40$ MeV, the mass-asymmetric fission events are
dominant. The experimental data of the mass distribution of
fission events for the QF process is denoted by the thick black
line in Fig.~10(a) \cite{mate02}. In the calculation, the peaks
located at $A\sim48$ and 244 correspond to a reaction similar to
the DIC process. That is to say, due to the low incident energy,
the colliding system reseparates like a projectile and a target.
The trajectory cannot overcome the first ridge on the $z-\alpha$
plane, and goes back to the $+z$ direction while maintaining the
initial mass asymmetry $\alpha$.

With increasing incident energy, the trajectory overcomes the
first ridge on the $z-\alpha$ plane and the events of the DIC
process decrease. Mass-symmetric fission events become dominant.
In Figs.~10(b) and (c), the peaks at $A\sim 80, 210$ grow. These
peaks originate from the QF process which goes down the valley
leading to the Pb like fragment, after overcoming the first ridge
on the $z-\alpha$ plane.
The QF process, which progresses to mass-asymmetric fission
events, decreases as a whole whereas the FF process increases when
incident energy increases.


\begin{figure}
\centerline{
\includegraphics[height=.51\textheight]{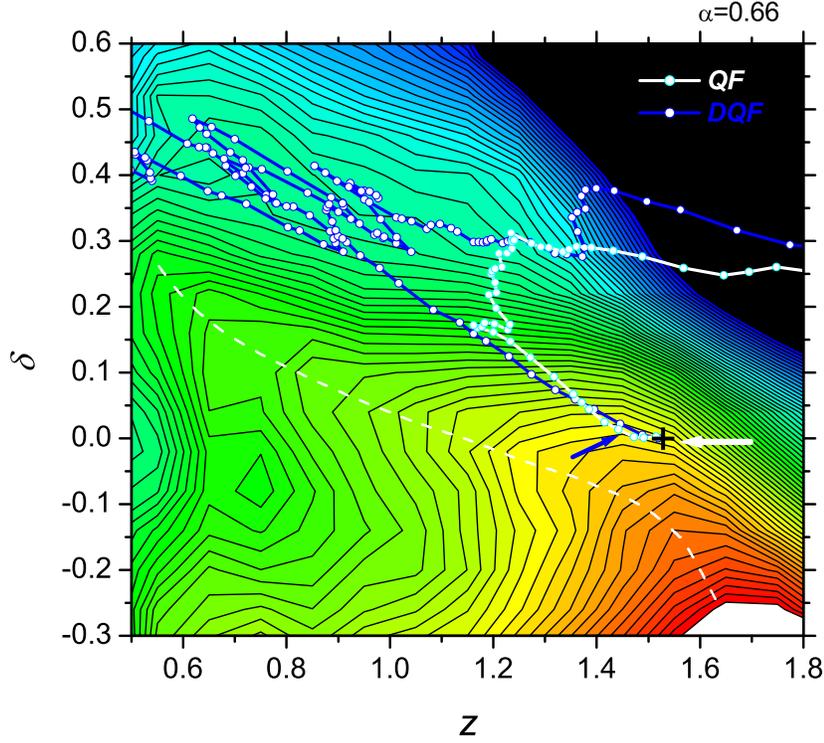}}
  \caption{Samples of the trajectory near point of contact projected onto
  $z-\delta$ $(\alpha=0.66)$ plane in reaction $^{48}$Ca+$^{244}$Pu
  at $E^{*}=40$ MeV. The QF and DQF processes are denoted by the
  white and blue lines, respectively. Symbols are given in the text.}
\end{figure}


\begin{figure}
\centerline{
\includegraphics[height=.51\textheight]{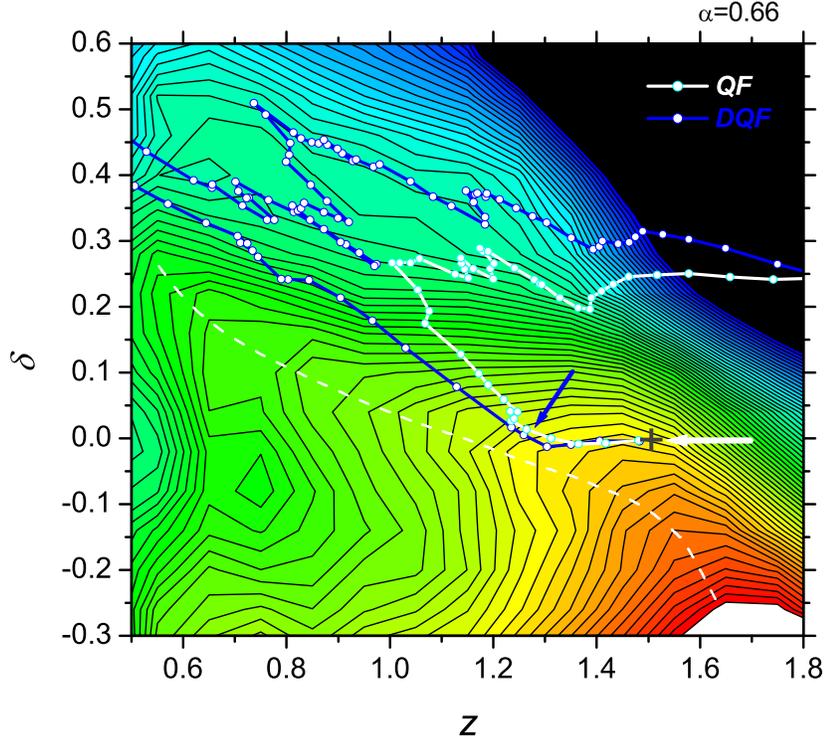}}
  \caption{Samples of the trajectory near point of contact projected onto
  $z-\delta$ $(\alpha=0.66)$ plane in the reaction $^{48}$Ca+$^{244}$Pu
  at $E^{*}=60$ MeV. The QF and DQF processes are denoted by the
  white and blue lines, respectively. Symbols are given in the text.}
\end{figure}

Next, we discuss further the critical stage. Figure~7 shows the
samples of the time evolutions of the kinetic energy of the
relative motion in the reaction $^{48}$Ca+$^{244}$Pu at $E^{*}=40,
50$ and 60 MeV, which are denoted by the dashed, solid and
dashed-dotted lines, respectively. We can see that the kinetic
energy in the incident channel dissipates within $t=3.0\times
10^{-22}$ sec independent of the incident energy. We denote this
dissipation time as $t_{diss}$. After this time, the trajectory is
controlled by random force and the landscape of the potential
energy surface.


Figures~11 and 12 show the trajectory near the point of contact in
the reaction $^{48}$Ca+$^{244}$Pu at $E^{*}=40$ and 60 MeV,
respectively. The QF and DQF processes are denoted by the white
and blue lines, respectively. The white arrow indicates the
injection direction and the point of contact is marked by +. By
investigating the trajectory of the FF process at $E^{*}=40$ and
60 MeV, we found that the trajectory of the FF process follows
almost the same path as that for $E^{*}=50$ MeV. In contrast, the
trajectories of the DQF and QF processes show the characteristic
behaviors at each incident energy. In Figs.~11 and 12, the
interval between adjacent circles on the trajectory corresponds to
$6.58\times 10^{-23}$ sec, in the same manner as in Fig.~8.


At $E^{*}=40$ MeV, because of the low incident energy, the initial
trajectory speed is low. Until $t=t_{diss}$, the trajectory leaves
only from near the point of contact. The position at $t_{diss}$ is
denoted by the blue arrows in Fig.~11. As a result, the trajectory
travels down the steep slope in the $+\delta$ direction. Due to
random force applied during the decent down the slope, the
trajectory is separated into the DQF and QF processes.

In the case of $E^{*}=60$ MeV, because of the high incident
energy, the initial trajectory speed is high, and all trajectories
advance toward the second ridge in the $-z$ direction. As shown in
Fig.~12, the trajectory approaches to the second ridge at
$t_{diss}$, as denoted by the blue arrows.  After this point, the
trajectory is controlled by random force and the landscape of the
potential energy surface. The trajectory that cannot overcome the
second ridge descends the steep slope in the $+\delta$ direction.
It also is separated into the DQF and QF processes by random force
during the descent down the slope.



We have discussed the mechanism of the dynamical process in the
case of zero angular momentum. When the system has an angular
momentum, the potential landscape changes. That means the relation
between the ridge line and the point at $t_{diss}$ changes. As
discussed above, the behavior of the trajectory mainly is
controlled by this relation and random force. This main mechanism
can be applied to the any angular momentum cases.







\section{Summary}

The fusion-fission process in the superheavy-mass region was
studied on the basis of fluctuation-dissipation dynamics. The
trajectory calculation with friction has been performed.

In our previous work \cite{ari04}, we have compared our
calculation with experimental data of the mass distribution of
fission fragments and the cross section derived by counting mass
symmetric fission events. On the basis of our previous studies, we
investigated why the trajectory chooses a different path
corresponding to the different process. By analyzing the time
evolution of the trajectory, we clarify the mechanism of whole
fusion-fission dynamics. To find the condition to follow the
fusion path, it is strongly related to attempts to synthesize new
superheavy elements.

In the reaction $^{48}$Ca+$^{208}$Pb at $E^{*}=50$ MeV, the
fusion-fission process is dominant. The mechanism of this FF
process was clarified. The trajectory of the FF process is trapped
in the subpocket located at the intermediate deformed system, and
then moves to the main pocket located at the spherical nucleus. It
is found that trapping in the subpocket prevents the trajectory
from going to the fission region. We also investigated the time
evolution of the nuclear shape. We found that mass asymmetry is
rapidly relaxed when the trajectory moves quickly from the
subpocket to the main pocket.

In the reaction $^{48}$Ca+$^{244}$Pu, the dynamical process is
classified into the FF, DQF and QF processes. We investigated the
origin of each path by analyzing the time evolution of each
parameter. Under the assumption that the kinetic energy does not
dissipate during the approaching process, the time up to $t=0.5
\times 10^{-21}$ sec, which is called the critical stage, is very
important for separating each path. It roughly corresponds to the
time $t_{diss}$ at which all the kinetic energy dissipates. During
this critical stage, the point reached by the trajectory governs
the process after this stage. Then, the behavior of the trajectory
is controlled by random force and the landscape of the potential
energy surface. The trajectory of the FF process overcomes the
second ridge line on the $z-\delta$ plane, when the random force
applied in the direction of $-\delta$ and almost perpendicular to
the ridge. This is the key point for the FF process. The
difference between the origins of the DQF and QF processes was
also discussed.

The time evolution of the nuclear shape for each process was
discussed. The critical point of the difference between the QDF
and QF processes is the neck shape of the nucleus at the end of
the critical stage. While the nuclear shape in the QF process has
a neck at this time, the shape in the DQF process does not,
because of the large value of $\delta$ and the small value of $z$.
In the latter case, it is easy to form a mononucleus and thereby
easy to relax the mass-asymmetric parameter. Consequently, we can
see the mass-symmetric fission events in the DQF process.

From the investigation of the incident energy dependence of the
dynamical process, we can say that the position of the trajectory
at $t_{diss}$ depends on the incident energy and the position
decides the way of future dynamical path separation.

We believe that this study yields an effective method of
investigating the whole fusion-fission process in the
superheavy-mass region. For the synthesis of superheavy elements,
it is very important to know the mechanism of the fusion process
precisely.

The authors are grateful to Professor Yu.~Ts.~Oganessian,
Professor M.G.~Itkis, Professor V.I.~Zagrebaev and Professor
F.~Hanappe for their helpful suggestions and valuable discussion
throughout the present work. The authors thank Dr. S.~Yamaji and
his collaborators, who developed the calculation code for
potential energy with two-center parameterization. This work has
been in part supported by INTAS projects 03-01-6417.


\bibliographystyle{aipproc}   

\bibliography{sample}

\IfFileExists{\jobname.bbl}{}
 {\typeout{}
  \typeout{******************************************}
  \typeout{** Please run "bibtex \jobname" to optain}
  \typeout{** the bibliography and then re-run LaTeX}
  \typeout{** twice to fix the references!}
  \typeout{******************************************}
  \typeout{}
 }

\end{document}